\documentclass[namedreferences]{solarphysics}
\usepackage[optionalrh,solaromanenum,natbib]{spr-sola-addons} 
\usepackage[dvips]{graphicx}                    
\usepackage{amssymb}                    
\usepackage{color}                       
\usepackage{url}                         
\usepackage[utf8]{inputenc}

\definecolor{grey}{rgb}{0.5,0.5,0.5}

\begin{document}
\begin{article}
\begin{opening}

\title{Critical Analysis of a Hypothesis of the Planetary Tidal Influence on Solar Activity}
\author{S.~\surname{Poluianov}\sep
        I.~\surname{Usoskin}
       }
  \institute{University of Oulu, Finland
                     email: \url{stepan.poluianov@oulu.fi-a}
             }

\begin{abstract}
The present work is a critical revision of the hypothesis of the planetary tidal influence on 
solar activity published by Abreu \textit{et al.} 
(\textit{Astron. Astrophys.} \textbf{548}, A88, 2012; called A12 here).
A12 describes the hypothesis that planets can have an impact on the solar tachocline
and therefore on solar activity.
We checked the procedure and results of A12, namely the algorithm of planetary
tidal torque calculation and the wavelet coherence between torque and heliospheric modulation potential.
We found that the claimed peaks in long-period range of the torque spectrum are artefacts
caused by the calculation algorithm.
Also the statistical significance of the results of the wavelet coherence is found to be overestimated by an incorrect choice of the background assumption of red noise.
Using a more conservative non-parametric random-phase method, we found that the long-period coherence between planetary torque and heliospheric modulation potential becomes insignificant.
Thus we conclude that the considered hypothesis of planetary tidal influence on solar activity is not based on a solid ground. 
\end{abstract}

\end{opening}


\section{Introduction}
Regular observations of sunspots
started the subject of temporal variations of solar activity.
Since the discovery of the 11-year solar cycle by Heinrich Schwabe 
and its later confirmation 
by Rudolf Wolf, a question about possible causes of the Schwabe cycle and 
longer-period variations of solar activity is one
 of the key issues for solar physics.
A simple and intuitively easy-to-accept explanation would be a possible influence of planets with stable rotation periods on the Sun.
Of particular interest was Jupiter with its 10.86-year orbital period.
Many attempts have been performed since then to explain the variability of solar activity 
in this way \cite[\textit{e.g.,}][]{jose65,bigg67} without a great success though.
The present paradigm is that the solar variability is defined by the solar dynamo process driven solely by the dynamics of the convection zone \cite[\textit{e.g.,}][]{charbonneau10}.
However, the idea of a possible planetary influence on the solar activity is still discussed.
In this work we debate a recent paper by \cite{abreu12} (called A12 henceforth), where finding
 of a statistical relation between variations of the planetary tidal forces on the Sun and solar 
 activity is claimed for the last 9400 years.
If correct, this result would have far-reaching implications for forecasts of solar activity for the next
 hundreds and even thousands of years \citep{charbonneau13}.
Therefore, we focus on a critical analysis of the method and data used in A12 that led them to this important conclusion.

We first attempt to repeat precisely the recipe by A12 in the data analysis.
Next we critically review the obtained results and discuss possible artefacts.
Finally, we draw a conclusion on the robustness of the results.
\section{Planetary Torque Calculations}

\subsection{Original Algorithm of A12}
The core of the A12 work is related to the proposed effect that the planetary tidal forces
 can make upon the tachocline of the Sun (a thin layer between the convective and radiative zones).
This may lead, in the case of a non-spherical tachocline, to torque which is equal to the product
 of the tidal force and the heliocentric distance to the given point of the tachocline.
The torque is a vector and its projections on orthogonal axes are defined for the
{\it i}th planet by the following formulae:
 \begin{equation}
    \label{Nx}
    N_{x,i} = \frac{3}{5}G\rho m_i \frac {r_{y,i}r_{z,i}} {|\mathbf{r}_i|^5} [V_2(e^2-f^2)-V_1(b^2-c^2)],
  \end{equation}
  \begin{equation}
    \label{Ny}
    N_{y,i} = \frac{3}{5}G\rho m_i \frac {r_{z,i}r_{x,i}} {|\mathbf{r}_i|^5} [V_2(f^2-d^2)-V_1(c^2-a^2)],
  \end{equation}
  \begin{equation}
    \label{Nz}
    N_{z,i} = \frac{3}{5}G\rho m_i \frac {r_{x,i}r_{y,i}} {|\mathbf{r}_i|^5} [V_2(d^2-e^2)-V_1(a^2-b^2)],
  \end{equation}
where $G$ is the gravitational constant, $\rho$ is the mass density in the tachocline,
$m_i$ is the mass of the {\it i}th planet, $r_{x,i}$, $r_{y,i}$, $r_{z,i}$ are heliocentric 
coordinates of the {\it i}th planet, $|\mathbf{r}_i|$ is the distance from the {\it i}th planet 
to the centre of the Sun and $V_1$ and $V_2$ are volumes of the hypothetic internal and external ellipsoids, 
describing the shape of the tachocline, with semi-axes $a$,$b$,$c$ and $d$,$e$,$f$, respectively. 
The range of $i$ from 1 to 8 means planets from Mercury to Neptune (see details in A12).

We use the same source of the planetary coordinates as A12, \textit{viz.} the NASA Jet Propulsion Laboratory Ephemeris DE408 relative to the equatorial J2000 coordinate system (\url{http://ssd.jpl.nasa.gov/?ephemerides}).
We consider the period from 7440 BC till 1977 AD (ages from 9389 till -27 years BP, where BP stands for Before Present, \textit{i.e.} before 1950).
This period corresponds to the solar variability data used by A12.

The total torque is a vectorial sum of torques from each of the planets:
\begin{equation}
  \mathbf{N} = \sum_{i=1}^8 \mathbf N_i.
\end{equation}

The algorithm to compute the torque, according to A12, is performed in three steps (Jose Abreu, personal communication, 2013):
\begin{enumerate}
  \item calculation of daily values of the torque projections on the orthogonal axes $N_x$, $N_y$ and $N_z$;
  \item annual averaging of the daily torque values;
  \item calculation of the modulus of the torque vector.
\end{enumerate}

\subsection{Discretization Problem: Theory.}
Here we demonstrate that the algorithm used by A12 contains an internal problem: 
the averaging period (one year, see step (ii) above)
 coincides with the Earth' orbital period and exceeds orbital periods of the inner planets.
As a consequence, the annual average of the torque vector for the Earth is close to zero by definition,
 and it takes unpredictable values 
 for the inner planets.
 For example, all the deviations of the Earth's orbit from the perfect periodic planar circle would lead,
 for the annual averaging, to spurious power peaks in the low-frequency range of the spectrum.
According to the Nyquist--Shannon sampling theorem (also known as the Kotelnikov theorem), the continuous signal
 is unambiguously determined by discretization with frequency not less than double value of the maximum frequency 
 in the signal spectrum (Nyquist frequency) \cite[\textit{e.g.,}][]{lyons01}.
 The Nyquist frequency for the Mercury ephemeris (\textit{i.e.} coordinates) with the shortest orbital period of
 0.24 years is $2/0.24$~year$^{-1} = 8.333$~year$^{-1}$, which is the minimum sampling frequency for the Mercury ephemeris.
We note that, for the torque data, this value should be doubled because the product of coordinates like
 $r_x \cdot r_y$ in Equations (\ref{Nx})--(\ref{Nz}) means effective doubling of the orbital frequency.
Thus the Nyquist frequency for the Mercury-induced torque, as well as for the total torque, is
 $2\times8.333$~year$^{-1} = 16.667$~year$^{-1}$.

If a continuous signal is digitized with the sampling frequency lower than the Nyquist value, its spectrum
 is known to be distorted.
The "true" high frequency part of the spectrum does not disappear but gets "projected" into the low-frequency
 range of the spectrum.
This effect is known as aliasing \cite[\textit{e.g.,}][]{lyons01}.

\begin{figure}[p]
  \centering
    \includegraphics[width=1\textwidth]{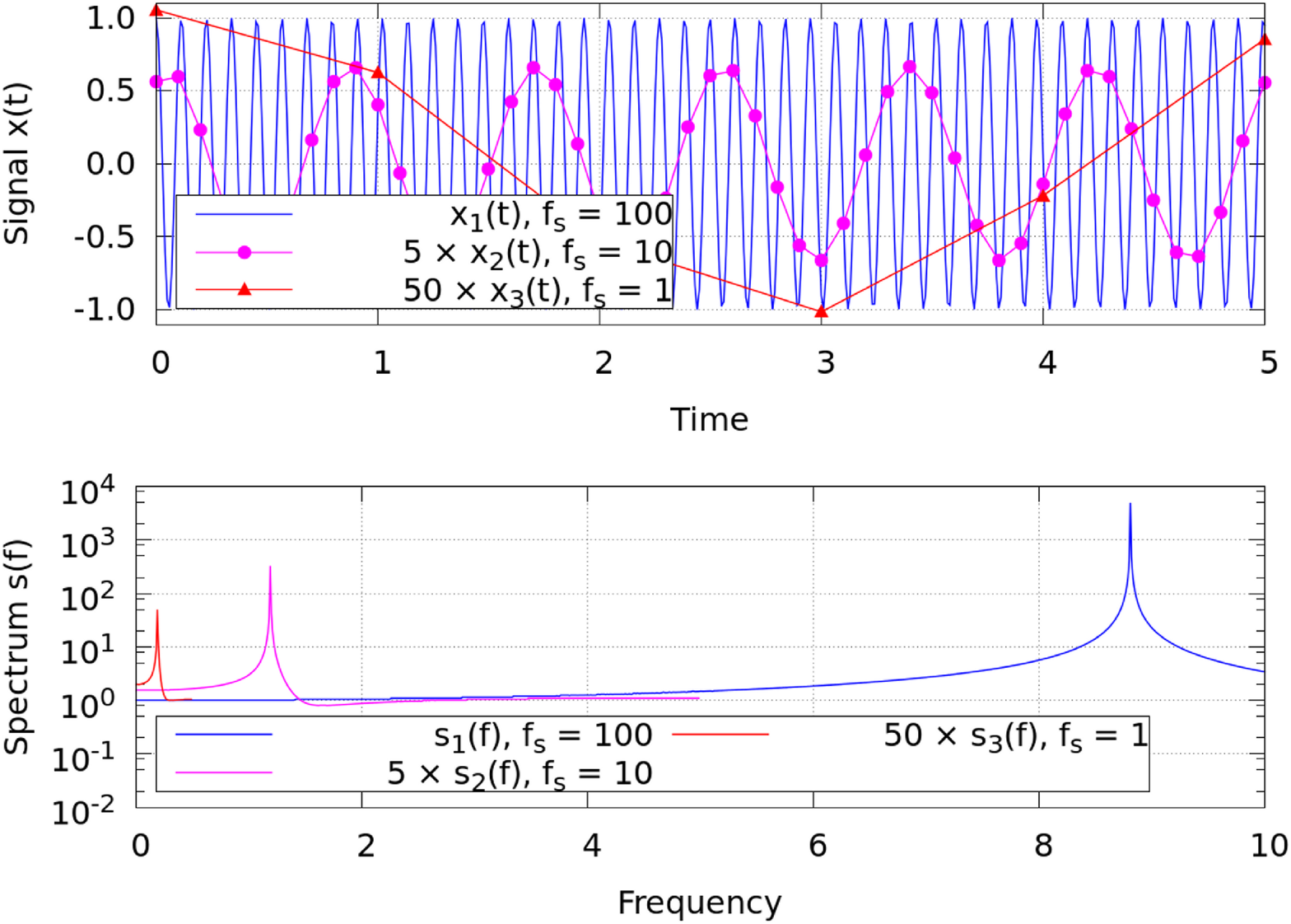}
    \caption{An illustration of the aliasing effect.
    The results of discretization of continuous harmonic signal $\cos (2\pi f_0 t)$ ($f_0=8.81$) by averaging with sampling frequencies $f_{\rm s}=$ 1, 10, and 100 (top panel) and their spectra (bottom panel).}
    \label{aliasing}
\end{figure}

The averaging of a signal as done in A12 (see step (ii) of the algorithm) unavoidably leads to this.
Figure~\ref{aliasing} illustrates an example of aliasing.
Let us assume a continuous harmonic signal with fixed frequency $f_0 = 8.81$.
The Nyquist frequency of the signal is $f_{\mathrm N} = 2f_0 = 17.62$.
The signal is sampled by averaging with three frequencies $f_{\mathrm s} = 1, 10$, and 100.
The resulting discrete signals and their spectra are shown on top and bottom panels, respectively.
The first two signals with $f_{\mathrm s} < f_{\mathrm N}$ are distorted and their spectral peaks are shifted
 from the true position of $f_0$ to frequencies 0.188 and 1.189.
The last one with $f_{\mathrm s} > f_{\mathrm N}$ does not have any aliasing distortion.
Its spectral peak stands at the frequency that is equal to $f_0$.

The raw ephemeris and torque data have the sampling frequency 1~day$^{-1}$~$\approx 365$~year$^{-1}$,
 which is much higher than the Nyquist frequency for all the planets, thus no aliasing is expected.
But averaging them to annual values (sampling frequency 1~year$^{-1}$) leads to the aliasing effect and distorts spectrum.
The aliasing effect can be crucial for the A12 analysis since they consider low-frequency (long-period) variations.
In the following subsection we perform a numerical experiment to check this.

\subsection{Discretization Problem: Numerical Check}

In order to check if the aliasing effect plays a role in the A12 analysis, we compute the planetary torque data,
 and consequently their power spectra, with different sampling frequencies: two frequencies lower and one much higher 
than the Nyquist frequency discussed above.
Since the aliasing effect is dependent on the sampling frequency, different sampling frequencies are expected
 to produce different aliasing distortions to the power spectrum.
On the other hand, the sufficiently high sampling frequency should yield an aliasing-free spectrum.
Therefore, we compute the primary torque series with the daily sampling and then re-sample it by averaging
 to the sampling frequencies $f_{\rm s} = 1$~year$^{-1}$, 10~year$^{-1}$, and 365.24~year$^{-1}$.

\begin{figure}[p]
  \centering
  \includegraphics[width=1\textwidth]{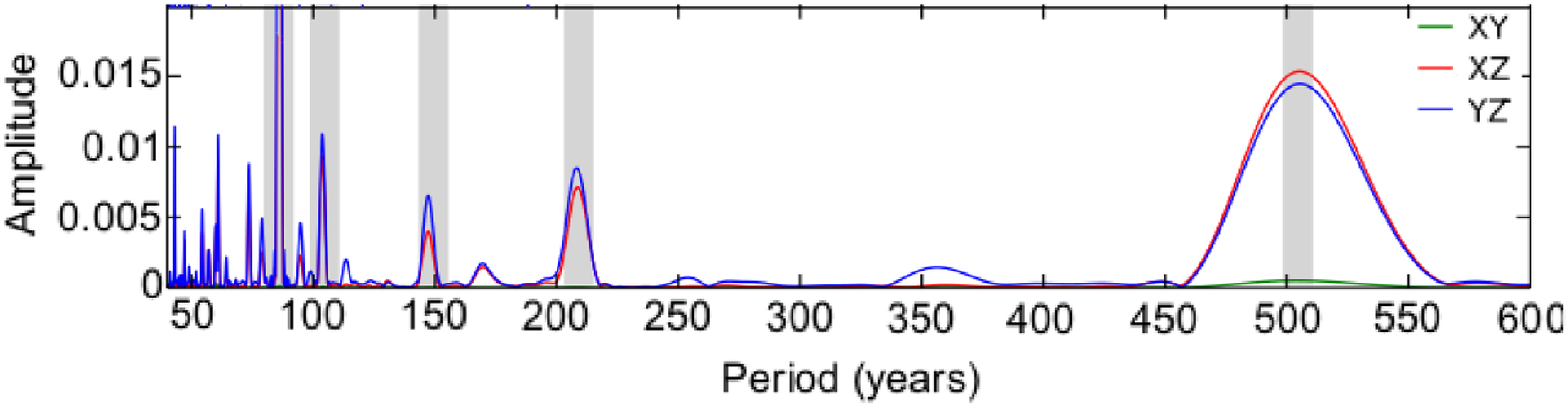}
  \caption{Original planetary torque spectra from A12 (Abreu {\it et al.}, 2012), calculated from the annually averaged data.
  Case 1 considered here corresponds to the "YZ" curve.
  The grey shaded areas denote the fundamental frequencies claimed by A12.}
  \label{abreu_spectra}
\end{figure}

Since the exact parameters of the tachocline ellipsoid are constant in time, and we focus on the periodicities
 here, we reduce Equations (\ref{Nx})--(\ref{Nz}) to
  \begin{equation}
    N_{x,i} = m_i \frac {r_{y,i}r_{z,i}} {|\mathbf{r}_i|^5},
  \end{equation}
  \begin{equation}
    N_{y,i} = m_i \frac {r_{z,i}r_{x,i}} {|\mathbf{r}_i|^5},
  \end{equation}
  \begin{equation}
    N_{z,i} = m_i \frac {r_{x,i}r_{y,i}} {|\mathbf{r}_i|^5}.
  \end{equation}
Only masses (constant) and coordinates (varying) of the planets enter these formulae.
The tidal effect is inversely proportional to the cube of the distance, thus the fraction
 ${m_i}/|\mathbf{r}_i|^3$ defines the relative contribution of the {\it i}th planet to the total torque.
Following the A12 recipe, we calculate the final power spectrum as a fast Fourier transformation (FFT)
 of the modulus of the planetary torque time series.
Here we deal mostly with the case~1 of A12 (the $N_x$ component is set to zero, the "YZ" component in
 Figure~\ref{abreu_spectra}), but note that the spectra for the other two cases of A12 ($N_y=0$ and $N_z=0$, 
 respectively) have similar frequency configurations.

\begin{figure}[p]
  \centering
  \includegraphics[width=1\textwidth]{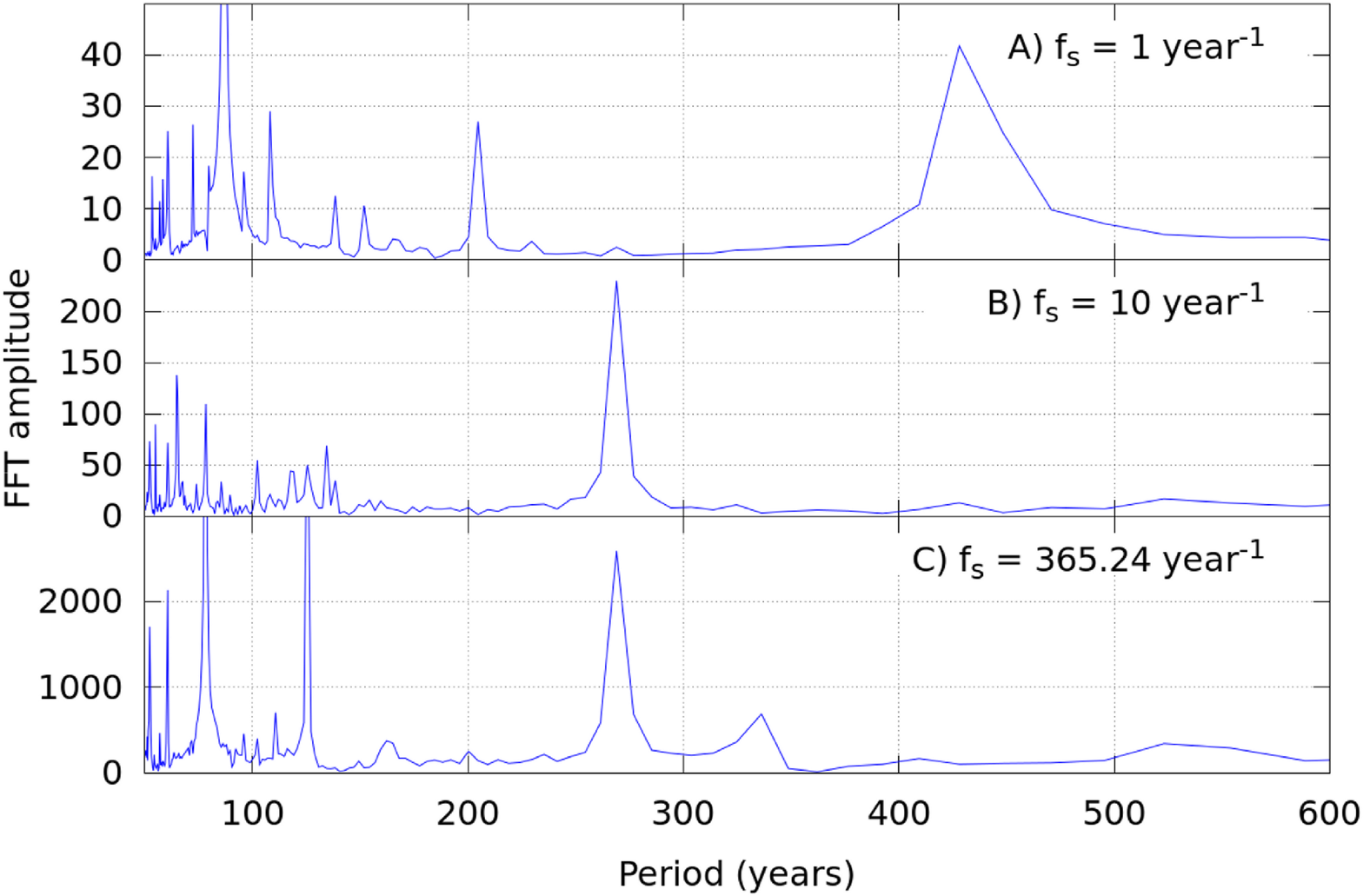}
  \caption{The planetary torque spectra computed here for the three sampling frequencies
   (see definition in the text): 1, 10, and 365.24 yr$^{-1}$ for panels A--C, respectively.}
  \label{spectra}
\end{figure}
Figure~\ref{abreu_spectra} shows the original spectra of the (modulus of) planetary torque for 
three cases from A12, who used annually averaged torque data.
The grey intervals indicate fundamental periodicities of 88, 104, 150, 208, and 506 years claimed by A12.

Figure~\ref{spectra} depicts three spectra of the same (modulus of) planetary torque as A12, but
 computed here with different sampling frequencies according to the algorithm by A12.
One can see (panel A) that the spectrum with $f_{\rm s}=1$~year$^{-1}$ (which corresponds to the
 case of A12) does show several peaks with periods identical or very close to those found by
 A12 (\textit{cf.} Figure~\ref{abreu_spectra}).
The spectrum contains several clearly defined spectral peaks at about 61, 72, 88, 108, $\approx 205$, and 430~years.
The only minor difference is that we found the 506-year peak, claimed by A12, at a shorter period ca. 430 years.
This may be related to the fact that the procedure is extremely sensitive in the long-period range, and a small difference in the detail of the FFT computation may lead to the "floating" peak.
This also illustrates that long periods are not robustly defined here.

Interestingly, the spectrum with the sampling frequency $f_{\rm s}=10$ year$^{-1}$ (panel B) is 
dramatically different from that for the annually sampled data.
The longer peaks dominating the long-period range in the annually sampled data disappear, while 
a new very strong peak appears at about 270 years.
In the shorter-period range, the pattern is much more noisy with numerous peaks at 
periods shorter than 140 years.
A number of peaks are found in the period range 100--140 years, strong peaks occur at 79 years and 65 years, and again several peaks between 50 and 60 years.
Such a noisy pattern is typical for the power migrating from high to low frequencies because of the aliasing effect.

Now we compute the power spectrum for the original data with the daily resolution
$f_{\rm s}=365.24$ year$^{-1}$ (panel C).
This spectrum does not have an aliasing distortion and is considered as the "true" reference spectrum.
It has several pronounced peaks at 270, 126, 79, 61, and 52 years.
It is important that none of them has a counterpart in the annually sampled data ($f_{\rm s}=1$ year$^{-1}$).
In fact, the spectrum computed from the annually averaged data has nothing in common with the "true" spectrum.
Meanwhile, for the $f_{\rm s}=10$ year$^{-1}$, some peaks remain (270 and 79 years) but change their amplitudes,
 others move in frequency or split.

This implies that the spectral peaks in the planetary torque series claimed by A12 are caused by an artefact of the applied method, \textit{viz.} the aliasing effect because of the annual averaging of the data before processing.

\section{Comparison of Solar Activity and Planetary Torque}
In the second part of this work we focus on an analysis of the wavelet-coherence calculations between the heliospheric modulation potential (as a tracer of solar activity) and planetary torque as done by A12.
We use the same data and method for the analysis.
We only apply a more appropriate procedure of the significance estimation of the obtained results as compared to A12.

\subsection{Computation of the coherence and its significance}

As an index of solar activity variations, A12 made use of the heliospheric modulation potential, which is a very convenient parameter to characterize solar modulation of cosmic rays \citep{usoskin_phi_05}.
Following the procedure by A12, we use the modulation potential reconstructed for the last millennia by \cite{steinhilber12} from cosmogenic radionuclides $^{10}$Be and $^{14}$C in natural archives such as ice cores and tree rings.
These data cover the age range from 9389 BP till present.
Here we use the original planetary torque series that is directly obtained from the authors of A12 (Jose Abreu, personal communication, 2013), not calculated in the first part of the present work.

We apply the wavelet-coherence method to estimate the relation between the modulation potential and the planetary torque data sets.
We use the core of Matlab-based package developed by \cite{grinsted04}, \textit{i.e.} the same as used by A12.
The method allows us to estimate the coherence
between two data sets both in time and frequency domains as well as the phase relation between the series.

We note that the original method to estimate statistical significance of the coherence is based on
red noise as the first-order autoregressive model AR(1) \citep{grinsted04}.
However, as discussed by \cite{usoskin_JASTP_06} and shown in the appendix, such a method may 
essentially overestimate the significance of strong peaks in the spectrum.
Instead, a non-parametric random-phase method by \cite{ebisuzaki97} 
should be used \citep[\textit{e.g.,}][]{sugihara12}.
The method is based on a Monte-Carlo estimate of the significance, with random mixing of the
original signal phases but keeping their power spectra.
A comparison of two methods is given in the appendix.
We have upgraded the original Matlab code by A. Grinsted et al. accordingly to accommodate this 
non-parametric method.
When evaluating the significance we have performed 1000 random-phase realizations for each series.

In addition to the full wavelet coherence, we use an integrated coherence spectrum, which is the average of the wavelet coherence over the time domain, excluding the cone of influence \citep{grinsted04}.
It is important to note that averaging is performed over complex values; thus this is a phase-relative operation.
The statistical significance of the integral coherence spectrum is calculated in the same way as for the full wavelet coherence using the non-parametric random-phase method.

\subsection{Results}
\begin{figure}[p]
  \centering
    \includegraphics[width=1\textwidth]{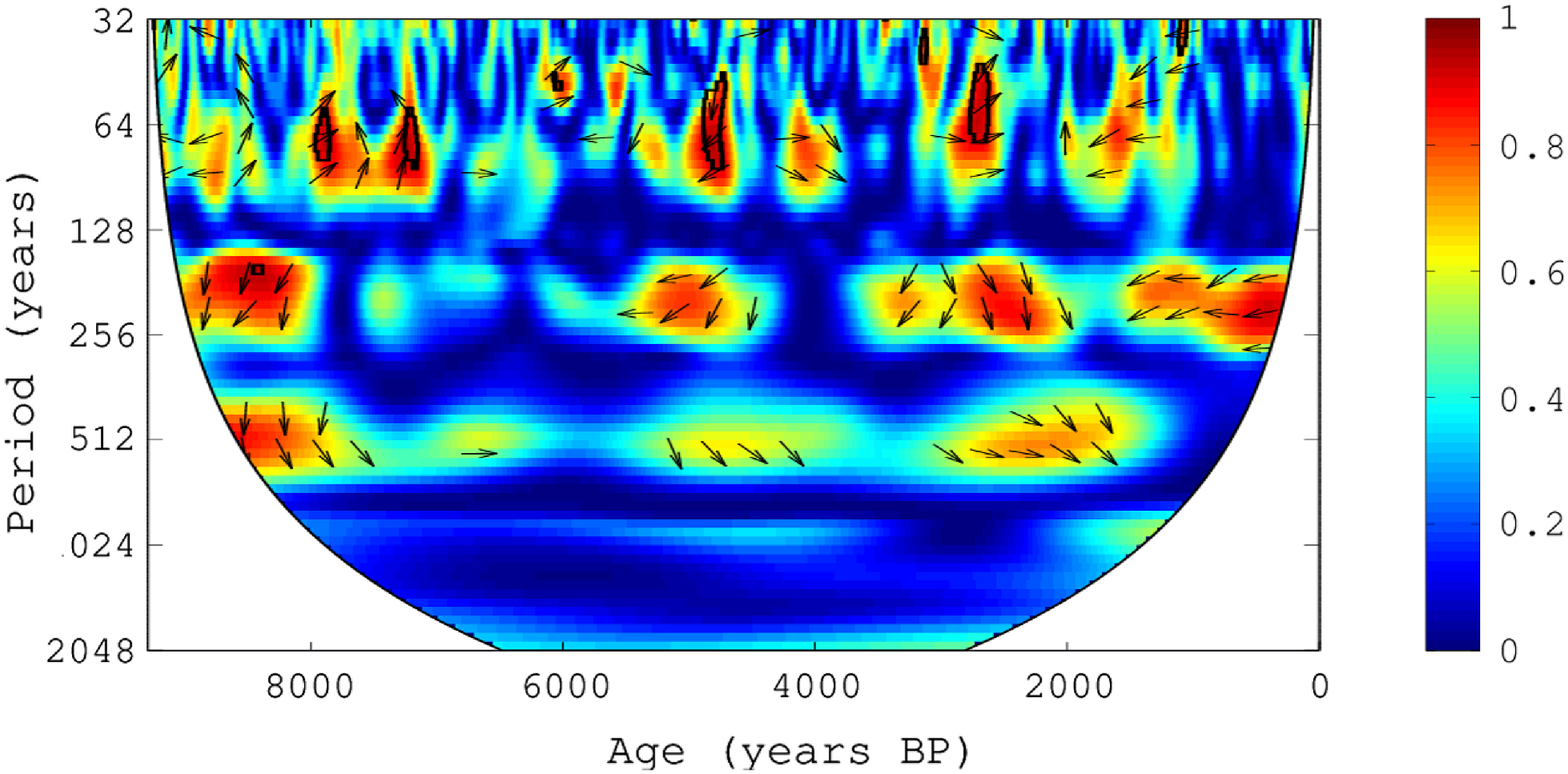}
    \caption{
    Wavelet coherence between the modulation potential and planetary torque series.
     Colours represent the coherence value from zero to one (see the colour bar).
     Arrows indicate phase difference between the two data sets (0 correspond to right-pointing arrows,
      90$^\circ$ up-pointing arrows, 180$^\circ$ left-pointing arrows).
      Black contours bound areas with the significance better than 5 \%.}
    \label{wtc}
\end{figure}

\begin{figure}[p]
  \centering
    \includegraphics[width=1\textwidth]{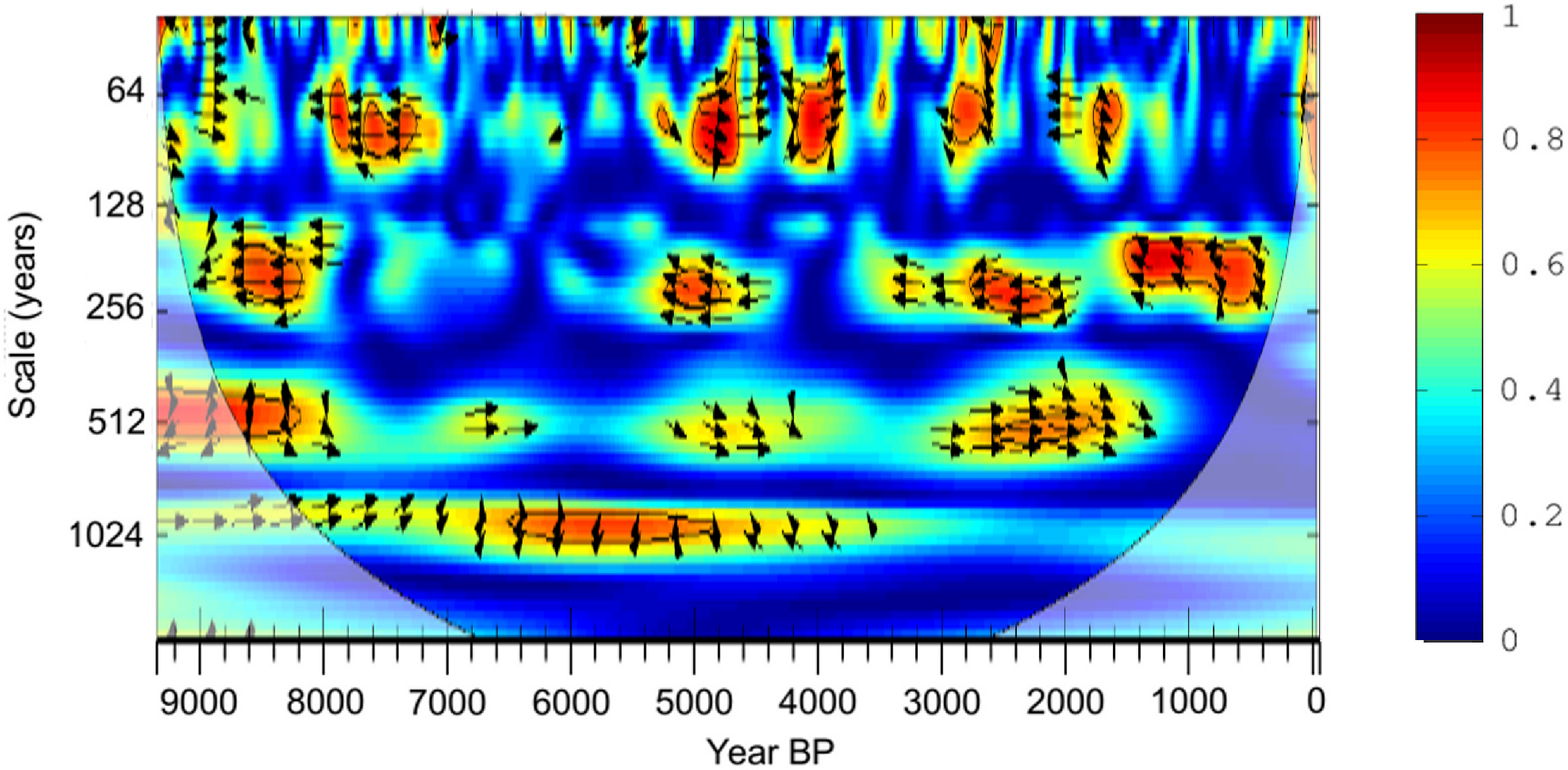}
    \caption{
    Wavelet coherence between the modulation potential and planetary torque series from the paper A12, case 1.
     Colours represent the coherence value from zero to one (see the colour bar).
     Arrows indicate phase difference between the two data sets (0 correspond to right-pointing arrows,
      90$^\circ$ up-pointing arrows, 180$^\circ$ left-pointing arrows).
      Black contours bound areas with the significance better than 5 \%.}
    \label{abreu_wtc}
\end{figure}

\begin{figure}[p]
  \centering
    \includegraphics[width=1\textwidth]{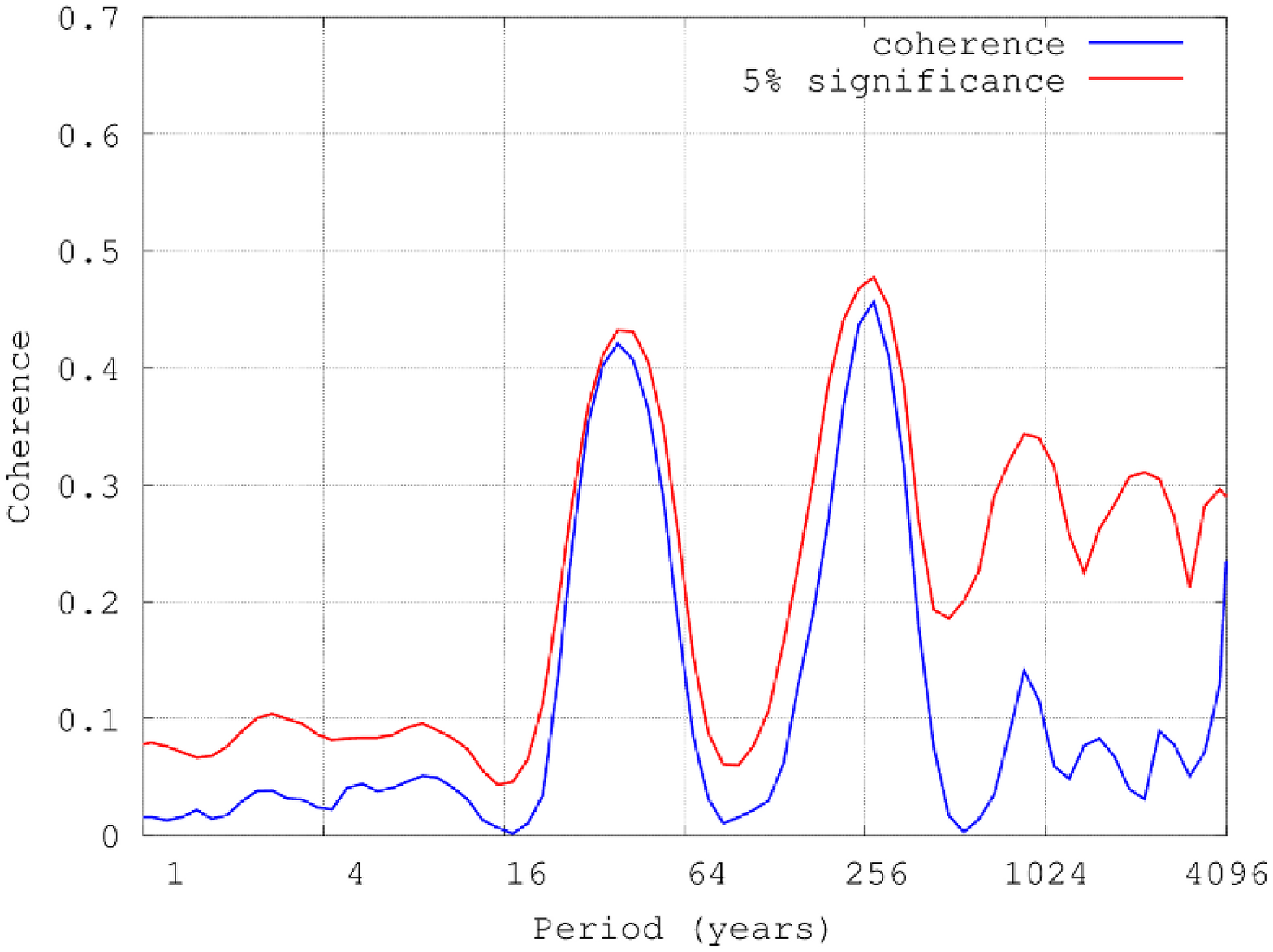}
    \caption{
    Time-integrated spectrum of the wavelet coherence between the modulation potential and
    planetary torque series.
    The dotted line denotes the 5 \% significance level.}
    \label{wtc2D}
\end{figure}

Figure \ref{wtc} shows the result of calculation of the wavelet coherence between the modulation potential and planetary torque series.
This plot is similar to the result of wavelet coherence from A12 (Figure~\ref{abreu_wtc}).
There are several lines of high coherence spots in the wavelet spectrum, although there are only a few  spots (at the period range of 60--70 years) which are significant at the 5\% significance level.
All other features are not significant.

The first line is at the period range of 60--100 years.
However, the relative phase between the series is out of order, with the arrows in the figure pointing to
different directions, resembling a random pattern.
In Figure~\ref{wtc2D} we show the integral wavelet-coherence spectrum, and one can see that the integral coherence is very low in this range of periods (60--100 years), because of the inconsistent phasing.
This suggests that the spotty coherence is not persistent.

Another sequence of coherence spots is observed at periods around 200 years.
The spots are repeated roughly every 2000 years with the duration of 300--500 years, which is in good agreement with the results of A12.
Because of the more or less stable phase, this leads to a strong peak (magnitude about 0.4) in the  integral coherence (Figure~\ref{wtc2D}).

The third sequence of the coherence spots lies in the range of periods of about 500 years.
The spots of about 500-year duration re-appear roughly 2500 years, in agreement with A12.
The relative phase is more or less stable around -90$^\circ$ (about 125-year delay), in contrast to the results of A12 who found a nearly in-phase relation.
This also lead to a pronounced peak (at the magnitude of about 0.4) in the integral coherence (Figure~\ref{wtc2D}).

Figure \ref{wtc2D} summarizes the integral wavelet coherence between the two series.
Only two pronounced peaks are present, around 210 and 500 years, respectively.
However, an estimate of the significance, made by the non-parametric random-phase method described above,
implies that even these two peaks are not statistically significant at the 5\% level (dotted curve).
The 210-year peak is barely significant at the 10\% level, while the 500-year peak is insignificant.
This can be understood so that a periodic signal does show some level of coherence even with a pure noise, by means of non-zero cross-spectrum (see the appendix).

Thus, we found that the coherence between the solar activity and the planetary torque series is not statistically significant and may be an artefact of combining a periodic and a noisy series.

\section{Conclusions}
We analysed the procedure of planetary torque calculations from the paper by \cite{abreu12}
 and found that their results can be be affected by an effect of the aliasing distortion of the torque spectrum.
We provided torque calculations with different sampling frequencies and found that the spectral peaks
 claimed by A12 are likely artefacts of the spectral distortion and do not have physical meaning.
Then we repeated the analysis by A12 of the relation between heliospheric modulation potential and the planetary torque.
We showed that the results of \cite{abreu12} are not statistically significant.
Thus, the proposed hypothesis of planetary influence on solar activity is not based on solid empirical evidence.

\section*{Acknowledgements}
The authors acknowledge Jose Abreu and J\"{u}rg Beer for providing the original data, details of the algorithm of planetary torque computation, and for the stimulating discussion.

\section*{Appendix: On the significance of Coherence between Narrow- and Wide-band Signals}
It is important to note that one must be careful when computing the coherence between 
narrow- and wide-band signals.
an incorrectly assessed significance of the coherence can produce false physical conclusions.

\begin{figure}[p]
  \centering
    \includegraphics[width=1\textwidth]{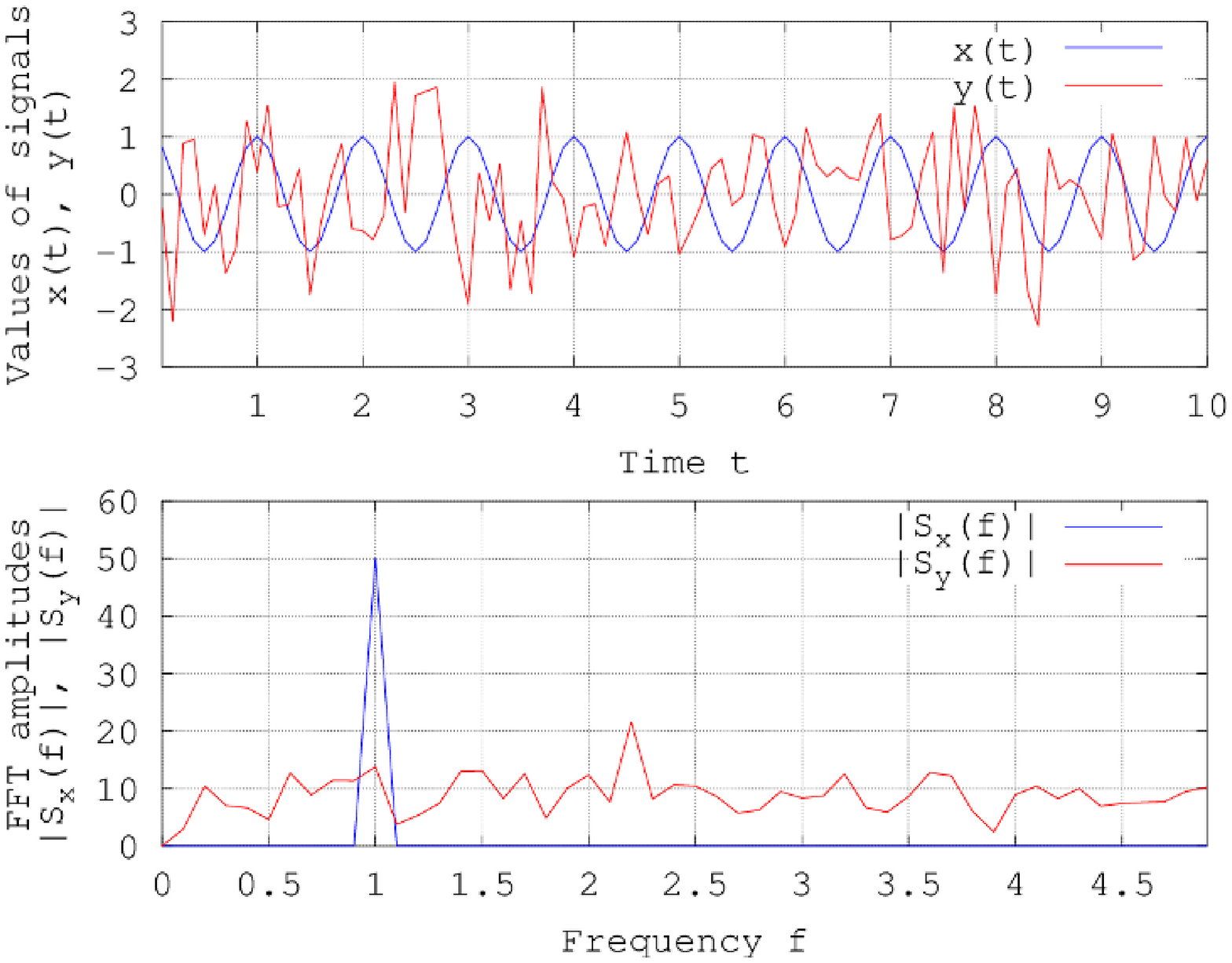}
    \caption{
    An illustration of computation of coherence between narrow- and wide-band signals. 
    The considered signals $x(t)$ and $y(t)$ are given in the top panel and 
    their Fourier spectra $S_x(t)$ and $S_y(t)$ are in the bottom panel.
    }
    \label{sigNspec}
\end{figure}

We illustrate it by a simple and clear numerical example.
Let us generate two independent and non-coherent narrow- and wide-band signals and call them $x(t)$ and $y(t)$, respectively.
The first one is a purely harmonic signal $x(t)=\cos(2\pi f_0 t)$, where 
$f_0$ is the frequency and $t$ is the time.
The second one $y(t)$ is white noise with normal distribution, zero mean and unity standard deviation.
The signals and their Fourier spectra are shown in Figure~\ref{sigNspec}.

Now let us calculate the formal coherence of two signals by the following formula:
\begin{equation}
  \label{eq:coherence}
  C_{xy}=\frac{|S_{xy}|^2}{S_x S_y},
\end{equation}
where $S_{xy}$ is the cross-spectrum of $x(t)$ and $y(t)$, $S_x$ and $S_y$ are Fourier spectra of $x(t)$ and $y(t)$, respectively.
A cross-spectrum is defined as:
\begin{equation}
  \label{eq:cross-spec}
  S_{xy}=S_x S_y^*,
\end{equation}
where the symbol "$^*$" means a complex conjugate.
The product given by Equation (\ref{eq:cross-spec}) extracts a narrow frequency range from the wide-band signal $y(t)$ by the narrow-band signal $x(t)$.
The coherence defined by the cross-spectrum [Equation (\ref{eq:coherence})] has non-zero values only near the frequency of the narrow-band signal $x(t)$.
Thus, a formal non-zero coherence exists between the two unrelated signals.

The described feature exists not only for Fourier analysis but for other kinds of spectra including wavelet analysis as well.

\begin{figure}[p]
  \centering
    \includegraphics[width=1\textwidth]{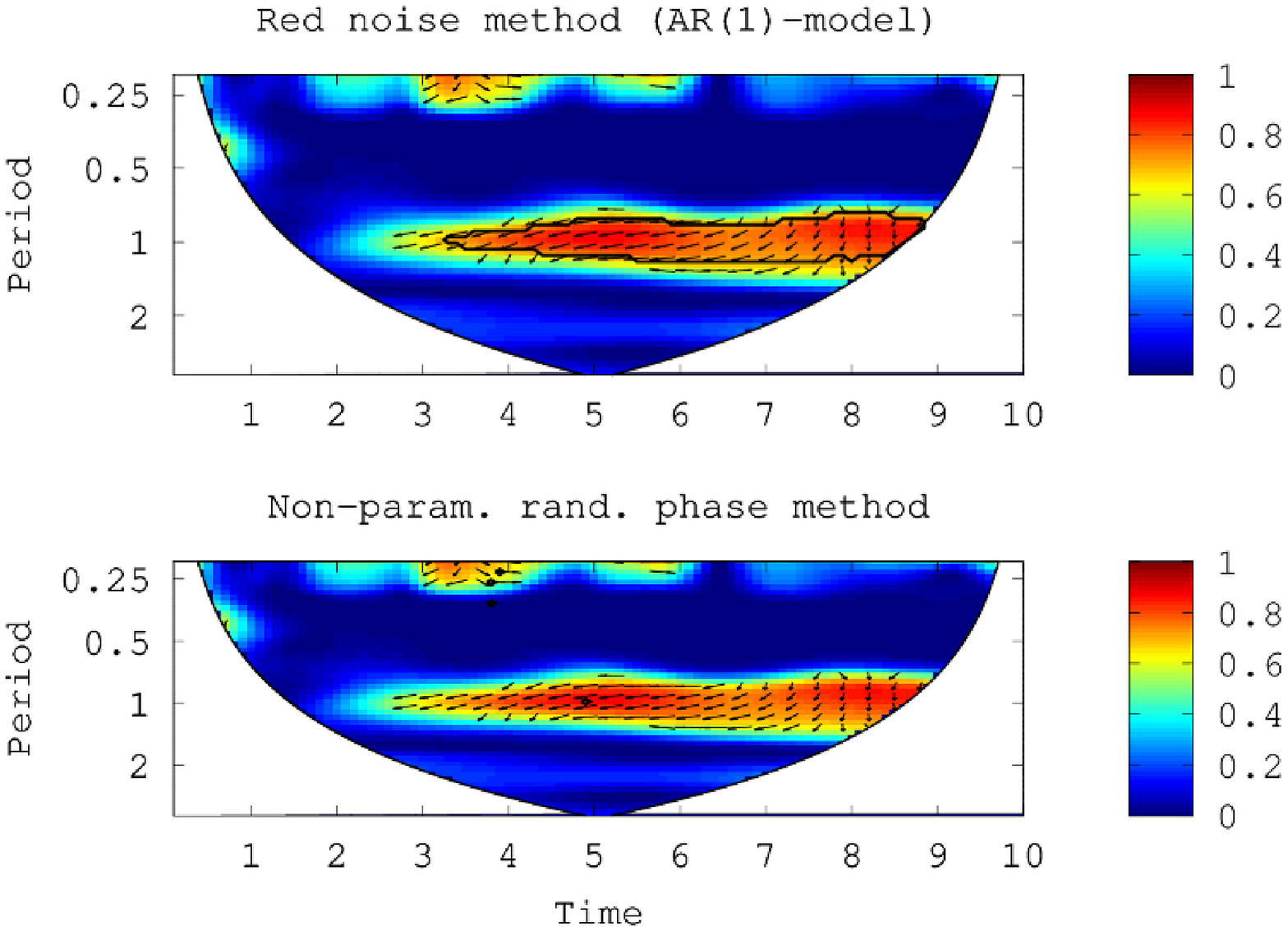}
    \caption{
    An illustration of computation of coherence between narrow- and wide-band signals. 
    The wavelet coherence between signals $x(t)$ and $y(t)$ with 
    statistical significance is calculated by two different methods: 
    autoregressive model AR(1) (red noise, in the top panel)
    and non-parametric random-phase method (in the bottom panel). 
    Significance is shown as black contours. 
    }
    \label{sig_wtc}
\end{figure}

The result of calculation of the wavelet-coherence for the two synthetic signals is presented in Figure~\ref{sig_wtc}.
The contours that indicate statistical significance areas are based on the autoregressive model AR(1) (red noise)
in the top panel and on the non-parametric random-phase method in the bottom panel.
There is some coherence between the signals $x(t)$ and $y(t)$.
Since two signals are non-coherent by definition, the computed coherence should not be statistically significant.
However one can see that the AR(1)-method estimates the coherence as significant.
The non-parametric random-phase method estimates coherence between $x(t)$ and $y(t)$ as insignificant for the same conditions.
It corresponds to the initial properties of the signals.

The present illustration is close to the case of computation of coherence between heliospheric modulation potential and planetary torque.
The former one has wide-band spectrum while the spectrum of the latter consists of a few narrow peaks.
It leads to the described effect and explains the derived coherence spots and their statistical insignificance in Figure~\ref{wtc}.


\end{article}
\end{document}